\documentclass[11pt]{article}
\usepackage{epsfig}

\newcommand{\const}{\,{\rm const}\,}

\def\be{\begin{equation}}
\def\ee{\end{equation}}
\def\bea{\begin{eqnarray}}
\def\eea{\end{eqnarray}}

\title{Do supersymmetric anti-de Sitter black rings exist?}

\author{Hari K. Kunduri \\School of Physics and Astronomy, University of Nottingham, NG7 2RD, UK \\h.k.kunduri@nottingham.ac.uk \\ \\ James Lucietti \\ Centre for Particle Theory, Department of Mathematical Sciences, \\ University of Durham, South Road, Durham, DH1 3LE, UK\\ james.lucietti@durham.ac.uk \\  \\Harvey S. Reall \\ School of Physics and Astronomy, University of Nottingham, NG7 2RD, UK \\harvey.reall@nottingham.ac.uk \\ \\ DCPT-06/35}
\date{1 December, 2006}

\begin{document}

\maketitle

\begin{abstract}
We determine the most general near-horizon geometry of a
supersymmetric, asymptotically anti-de Sitter, black hole solution
of five-dimensional minimal gauged supergravity that admits two
rotational symmetries. The near-horizon geometry is that of the
supersymmetric, topologically spherical, black hole solution of
Chong {\it et al}. This proves that regular supersymmetric anti-de Sitter
black rings with two rotational symmetries do not exist in minimal supergravity. However, we do find a solution corresponding to the near-horizon geometry
of a supersymmetric black ring held in equilibrium by a conical
singularity, which suggests that nonsupersymmetric anti-de Sitter
black rings may exist but cannot be "balanced" in the
supersymmetric limit.
\end{abstract}

\section{Introduction}

Supersymmetric black hole solutions of five-dimensional gauged
supergravity have been known for a few
years~\cite{GR1,GR2,Chong:2005,Chong:2005b,KLR}. According to the
AdS/CFT correspondence \cite{adscft}, these black holes should correspond to
$1/16$ BPS states of ${\cal N}=4$ $SU(N)$ super Yang-Mills theory
on $R \times S^3$ or, equivalently, gauge-invariant local
operators of this theory on $R^4$. A puzzle arises~\cite{KMMR}
because BPS operators carry five independent conserved charges: 
their spins $J_1,J_2$ and R-charges $Q_1,Q_2,Q_3$ (where
$(Q_1,Q_2,Q_3)$ is a weight vector of the $SU(4)$ R-symmetry
group), whereas the most general known asymptotically $AdS_5
\times S^5$ supersymmetric black hole solution has only 4
independent conserved charges~\cite{KLR}.

There are two ways that this puzzle could be resolved. (i) We
already know the most general supersymmetric black hole solution.
Although generic 1/16-BPS operators have 5 independent charges, 
only a subset of these operators gives a large enough
entropy to correspond to a macroscopic event horizon and this
subset has only 4 independent charges. (Maybe because of finite coupling effects in the CFT.) This possibility has been explored in \cite{berkooz}. (ii) There exists a
5-parameter supersymmetric black hole solution that remains to be
discovered.

If (ii) is correct, there appears to be some tension with what is
known about black holes in {\it minimal} gauged supergravity. One
can truncate to this theory by setting $Q_1=Q_2=Q_3=Q$. This
theory admits a 4-parameter non-supersymmetric black hole
solution~\cite{Chong:2005}. The 4 parameters are the 4 conserved
charges of this theory, namely $J_1,J_2,Q$ and the mass $M$. One
might expect this to be the most general black hole of spherical
topology. In the supersymmetric limit, one loses two parameters:
supersymmetric black holes are parameterized by $J_1$ and $J_2$.
Returning to general (unequal) $Q_i$, we would expect the most
general nonsupersymmetric black hole to be parameterized by the 6 conserved charges
$J_1,J_2,Q_1,Q_2,Q_3,M$ and losing 2 parameters in the supersymmetric
limit would take one to the 4-parameter solution of \cite{KLR}.
There does not seem to be any room for an additional parameter.

This objection rests on the assumption that black holes should be
characterized by their conserved charges. Even in four dimensions,
there is no uniqueness theorem for asymptotically AdS black holes,
so maybe this assumption is incorrect even for topologically
spherical AdS black holes. Furthermore, we know that this
assumption is violated by black rings \cite{vacring} in five asymptotically flat
dimensions, which can require nonconserved charges to specify them fully \cite{dipole}.  It is natural to guess that the same
is true in AdS. So perhaps the black holes of (ii) are supersymmetric
AdS black rings.\footnote{Other possibilities are that the black holes of (ii) do not admit a five-dimensional interpretation or that they involve non-abelian gauge fields.}

The goal of this paper is to classify supersymmetric black holes
in five-dimensional gauged supergravity. Unfortunately, finding
rotating black hole solutions is hard, even with supersymmetry. We
shall therefore adopt the approach initiated in~\cite{Reall:NH} of
classifying {\it near-horizon geometries} of supersymmetric black
holes\footnote{See~\cite{Davis, Her, SSS} for other recent work on the near-horizon geometry of supersymmetric AdS black holes.}. Obviously the existence of a near-horizon geometry with
certain properties cannot be taken as a proof of the existence of
a full black hole solution with those properties but this approach
{\it can} be used to rule out certain types of solution. For
example, if we find that near-horizon geometries with horizon
topology $S^1 \times S^2$ are not possible then that would exclude
supersymmetric black rings.

In minimal five-dimensional gauged supergravity, a classification
of near-horizon geometries was attempted in \cite{GR1}. However,
the resulting equations proved too difficult to solve in full
generality without additional assumptions. (This is in contrast
with the ungauged theory, for which a full classification is
possible~\cite{Reall:NH}.) The assumptions made in \cite{GR1} were
too restrictive to encompass the solutions of~\cite{Chong:2005}.
In this paper, we return to the equations of \cite{GR1} and solve them by making a
weaker assumption, satisfied by all known five-dimensional black
hole solutions, whether asymptotically AdS or asymptotically flat,
including black rings~\cite{vacring,dipole,bpsring}. Our assumption is
the existence of two rotational symmetries.\footnote{Note that the
"stationary implies axisymmetric" theorem~\cite{HIW:2005} for
black holes only guarantees the existence of a single rotational
symmetry. Furthermore it does not apply to supersymmetric black
holes. However, one might expect the conclusion to be valid for such solutions since one can obtain them as limits of non-supersymmetric black
holes.}

Our result is that a supersymmetric asymptotically AdS black hole
satisfying this assumption must have a near-horizon geometry
locally isometric to that of the known supersymmetric black hole
solution of~\cite{Chong:2005}. Hence either supersymmetric AdS
black rings do not exist, or they are rather unusual in that they do not admit two rotational symmetries, which would mean that they have
less symmetry than any known five-dimensional black hole solution.

Could there exist new supersymmetric black holes of spherical
topology, e.g. a 3-parameter generalization of the solution
of~\cite{Chong:2005}? We noted above that this seems unlikely but
we cannot exclude the possibility that there exists such a
solution with the same near-horizon geometry as the 2-parameter
solution of~\cite{Chong:2005}, i.e., that a parameter is lost in
the near-horizon limit (although we are not aware of any example
in which this happens).

Our analysis is local: we do not enforce spatial compactness of
the horizon until the end. This enables us to demonstrate why
supersymmetric AdS black rings with the same symmetries as
asymptotically flat ones do not exist. We do indeed obtain a
solution that corresponds to the near-horizon geometry of a black
hole with $S^1 \times S^2$ topology. It is a warped product of
$AdS_3$ and $S^2$. However, the $S^2$ always has a conical
singularity. This suggests that AdS black rings may exist, but
they cannot be "balanced" in the supersymmetric limit without the
presence of external forces corresponding to the conical
singularity. The conical singularity vanishes when one takes the
limit of vanishing cosmological constant, and one recovers the
$AdS_3 \times S^2$ geometry of an asymptotically flat
supersymmetric black ring~\cite{bpsring}.

Interestingly, when lifted to a solution of type IIB supergravity,
our singular black ring near-horizon geometry is {\it locally}
isometric to a solution obtained in~\cite{Gaunt1}. It was shown
in~\cite{Gaunt1} that this solution {\it can} be extended to a
globally regular metric in ten dimensions, but it does not appear
possible to reduce the resulting solution to five dimension so any
interpretation in terms of five dimensional black holes is lost.

This paper is organized as follows. We start by deriving a general constraint on black holes with two rotational symmetries. Then we review the results of \cite{GR1}, presenting the equations relevant to our analysis and deriving some general results. We solve these equations subject to the assumption of two rotational symmetries. There are only two interesting solutions and we show that these correspond to the near-horizon geometry of a supersymmetric black ring with a conical singularity and the near-horizon geometry of the solution of \cite{Chong:2005} respectively. 
Section 3 concludes. Some details of the analysis are relegated to appendices.

\section{Supersymmetric near-horizon geometries}

\subsection{General constraints}

The bosonic sector of minimal $D=5$ gauged supergravity is
Einstein-Maxwell theory with a negative cosmological constant and
a Chern-Simons term for the Maxwell field ${\cal F}$. The unique
maximally supersymmetric solution of this theory is $AdS_5$ with
vanishing gauge field \cite{GG}. We shall denote by $\ell$ the
radius of this $AdS_5$ solution.

Consider an asymptotically $AdS_5$ solution of this theory (not
necessarily supersymmetric) that is stationary and admits two
rotational symmetries, i.e., there is a $R \times U(1) \times
U(1)$ isometry group. Let $k,m_1,m_2$ denote the Killing fields
that generate time translations and rotations respectively. We
assume that these Killing fields commute and also leave the
Maxwell field invariant. Now we can use a standard argument from
the theory of stationary axisymmetric solutions \cite{exact}: the Bianchi
identity and the fact that the Lie derivatives of ${\cal F}$ along
the Killing fields vanish imply that $m_1^\mu m_2^\nu {\cal
F}_{\mu \nu}$ is constant. However, since the solution is
asymptotically $AdS_5$, we can find coordinates so that the
asymptotic metric is \be
 ds^2 \sim - \left( 1 + \frac{r^2}{\ell^2} \right) dt^2 + \left( 1 + \frac{r^2}{\ell^2} \right)^{-1} dr^2 + r^2 \left( d\theta^2 + \sin^2 \theta d\phi_1^2 + \cos^2 \theta d\phi_2^2 \right),
\ee
where $k = \partial/\partial t$, $m_i = \partial/\partial \phi_i$ and $0 \le \theta \le \pi/2$. Therefore, in the asymptotic region, $m_1$ and $m_2$ vanish at $\theta = 0, \pi/2$ respectively. It follows that
\be
\label{eqn:mdotF}
 m_1^\mu m_2^\nu {\cal F}_{\mu \nu} \equiv 0.
\ee The same argument shows that $k^\mu m_i^\nu {\cal F}_{\mu \nu}
\equiv 0$ although we shall not need this result.

\subsection{Near-horizon limit}

\label{sec:nheqs}

Given a supersymmetric black hole we can take a near-horizon limit
as explained in \cite{Reall:NH} to obtain a supersymmetric
near-horizon solution. We want to classify such solutions.
Necessary and sufficient conditions for a near-horizon geometry to
be a supersymmetric solution of this theory were worked out in
\cite{GR1} using the method introduced in \cite{Reall:NH}. One can
introduce Gaussian null coordinates on the horizon so that the
near-horizon metric is \cite{Reall:NH} \be \label{NHmet}
 ds^2 = -r^2 \Delta(x)^2 dv^2 + 2 dv dr + 2 r h_a(x) dv dx^a + g_{ab}(x) dx^a dx^b,
\ee where the horizon is at $r=0$, $\partial/\partial v$ is
Killing, $\Delta$ is non-negative, and the metric $g_{ab}$ is the
metric on a spatial cross-section of the horizon. We shall denote
this 3-manifold as $H$. Supersymmetry implies the following
\cite{GR1}:
\begin{itemize}
\item{
There exists a globally defined unit 1-form $Z$ on $H$:
\be
 g^{ab} Z_a Z_b = 1,
\ee
where $g^{ab}$ is the inverse of $g_{ab}$.
}
\item{
The (near-horizon) Maxwell field is
\be
\label{eqn:nhmaxwell}
 {\cal F} = \frac{\sqrt{3}}{2} \left( - dv \wedge d(r\Delta) - \star h - \frac{2}{\ell} \star Z \right),
\ee
where $\star$ is the Hodge dual on $H$. Hence the Bianchi identity is
\be
 \label{eqn:bianchi}
 d \star \left( h + \frac{2}{\ell} Z \right) = 0.
\ee
}
\item{
The following equations hold on $H$
\be
\label{eqn:dh}
 \star dh - d\Delta - \Delta h = \frac{6 \Delta}{\ell} Z,
\ee
\be
\label{eqn:gradZ}
 \nabla_a Z_b = - \frac{\Delta}{2} (\star Z)_{ab} + g_{ab} \left( h \cdot Z + \frac{3}{\ell} \right) - Z_a h_b - \frac{3}{\ell}{Z_a Z_b},
\ee
where $h \cdot Z \equiv g^{ab} h_a Z_b$. This implies that
\be
\label{eqn:dZ}
 dZ = -\Delta \star Z + h \wedge Z.
\ee

}
\item{
The Ricci tensor of $H$ is
\be
\label{eqn:ricci}
 R_{ab} = \left( \frac{\Delta^2}{2} + h \cdot h + \frac{4}{\ell} h \cdot Z \right) g_{ab} - h_a h_b - \nabla_{(a} h_{b)} - \frac{6}{\ell} h_{(a}Z_{b)} - \frac{6}{\ell^2} Z_a Z_b.
\ee
}
\end{itemize}
These equations are not the complete set derived in~\cite{GR1} but
they are the only ones we will need here. We have not been able to
solve these equations in full generality. However, we have
obtained some general results that is it convenient to record
here.

The near-horizon geometry is {\it static} if, and only if, the Killing field $V \equiv \partial/\partial v$ is hypersurface orthogonal, i.e., $V \wedge dV \equiv 0$.
The following lemma gives the conditions for this to occur:

\medskip

\noindent {\it Lemma 1.} The following conditions are equivalent:
(a) the near-horizon geometry is static, (b) $dh \equiv 0$, (c)
$\Delta \equiv 0$.

\noindent {\it Proof.} Assume (a). Then the $rab$ components of $V
\wedge dV \equiv 0$ give $dh \equiv 0$ so (a) implies (b).

Now assume (b). If $\Delta$ is nonzero then equation
(\ref{eqn:dh}) implies that $Z=-(\ell/6) ( h + d\Delta/\Delta)$
and hence $Z$ is closed. Equation (\ref{eqn:dZ}) then implies that
$\Delta Z \wedge \star Z=0$, hence $Z=0$, in contradiction with
the fact that $Z$ has unit norm. Hence we must have $\Delta=0$
everywhere. Hence (b) implies (c).

Finally assume (c). Equation (\ref{eqn:dh}) shows that $dh \equiv
0$. But $\Delta \equiv 0$ and $dh \equiv 0$ implies $V \wedge dV
\equiv 0$. Hence (c) implies (a).

\medskip

Our second lemma shows that $\Delta$ cannot vanish anywhere if the
near-horizon geometry is non-static.

\medskip

\noindent {\it Lemma 2.} If $\Delta$ vanishes at a point then $\Delta$ vanishes everywhere.

\noindent {\it Proof.} Take the divergence of (\ref{eqn:dh}) and use equations (\ref{eqn:bianchi}) and (\ref{eqn:gradZ}) to obtain
\be
\label{eqn:Deltaeq}
 \nabla^2 \Delta = - \left( h + \frac{6}{\ell} Z \right) \cdot \nabla \Delta - \frac{8}{\ell} \left( h \cdot Z + \frac{3}{\ell} \right) \Delta.
\ee
Assume $\Delta=0$ at $p$. Then, since $\Delta \ge 0$, $\Delta$ is at a (global) minimum so $d\Delta=0$ at $p$. It then follows from this equation that $\nabla^2 \Delta=0$ at $p$. However, since this is a minimum, the eigenvalues of the Hessian of $\Delta$ must be non-negative, and $\nabla^2 \Delta$ is the sum of the eigenvalues so they must all vanish. Hence the Hessian vanishes at $p$: $\partial_m \partial_n \Delta=0$ at $p$.

Assume inductively that $\Delta$ and its first $2n$ derivatives vanish at $p$. Then the Taylor expansion at $p$ begins
\be
 \Delta = \frac{1}{(2n+1)!} x^{i_1} \ldots x^{i_{2n+1}} \partial_{i_1} \ldots \partial_{i_{2n+1}} \Delta (p) + \ldots,
\ee
where $x^i$ are normal coordinates at $p$. Now $\Delta$ has to be non-negative but this leading term changes sign under $x \rightarrow -x$ and so we must require it to vanish. Therefore the $(2n+1)$th derivatives of $\Delta$ must vanish at $p$. Now the expansion of $\Delta$ is
\be
 \Delta = \frac{1}{(2n+2)!} x^{i_1} \ldots x^{i_{2n+2}} M_{i_1 \ldots i_{2n+2}} + \ldots,
\ee
where the symmetric tensor $M$ is the $(2n+2)$th derivative of $\Delta$. From our induction hypothesis and equation (\ref{eqn:Deltaeq}), we can obtain
\be
 \partial_{i_1} \ldots \partial_{i_{2n}} \nabla^2 \Delta (p)= 0
\ee
which implies
\be
 M_{i_1 \ldots i_{2n} jj} = 0.
\ee
This implies that the leading term in the expansion of $\Delta$ is harmonic in $R^3$. But we also need this term to be non-negative, which implies that it attains its minimum at the origin. Then from the maximum principle, it follows that this term must vanish everywhere, i.e., $M=0$. So we've shown that the first $(2n+2)$ derivatives of $\Delta$ vanish. It follows by induction, if $\Delta=0$ at $p$ then all derivatives of $\Delta$ vanish at $p$ hence $\Delta \equiv 0$ by analyticity\footnote{One might
object to the assumption of analyticity. However, none of the results in
this paper will rely on this lemma.}.

\medskip

In summary, these lemmas reveal that a static near-horizon
geometry has $\Delta \equiv 0$ and a non-static one has
$\Delta>0$. We end this section by noting that a static
near-horizon geometry can arise from a non-static black hole.
Indeed, this is what happens for supersymmetric black rings in
ungauged supergravity~\cite{bpsring}. We shall see that the same
appears to be true in gauged supergravity.

\subsection{Including the symmetries}

Consider a supersymmetric, asymptotically anti-de Sitter black
hole admitting two rotational Killing fields $m_1$ and $m_2$. The
near-horizon solution will inherit these symmetries. Hence we are
interested in classifying near-horizon solutions for which there
exist two commuting Killing vector fields $m_1$, $m_2$ on $H$ that
preserve $h$, $\Delta$ and the Maxwell field ${\cal F}$. It
follows from (\ref{eqn:nhmaxwell}) that the Killing fields must
also preserve $Z$. Furthermore, the near-horizon solution will
also inherit the condition (\ref{eqn:mdotF}).

We can choose local coordinates $x^a=(\rho,x^i)$ so that $\partial/\partial x^i$ are Killing, the metric on $H$ is
\be
 g_{ab} dx^a dx^b = d\rho^2 + \gamma_{ij}(\rho) dx^i dx^j,
\ee
and $\Delta$ and the components of $h$ are functions only of $\rho$. It is convenient to allow the Killing fields $\partial/\partial x^i$ to be arbitrary linear combinations of $m_1$ and $m_2$, i.e., $\partial/\partial x^i$ need not have closed orbits. We are then free to perform $GL(2,R)$ transformations on the coordinates $x^i$ to simplify our analysis. We shall enforce the fact that orbits of $m_1$ and $m_2$ must close once we have determined a local solution.

It is convenient to parameterize the components of $h$ as
\be
 h_i = \Gamma^{-1} \gamma_{ij} k^j, \qquad h_\rho = -\frac{\Gamma'}{\Gamma},
\ee
where $\Gamma(\rho)$ is positive and a prime denotes a derivative with respect to $\rho$.

The spatial components of the Maxwell field strength can be decomposed as
\be
 {\cal F}_{ab} dx^a \wedge dx^b =  \frac{\sqrt{3}}{2} B_i (\rho) d\rho \wedge dx^i,
\ee
where we have used equation (\ref{eqn:mdotF}) to deduce that the $ij$ components must vanish. Comparing with equation (\ref{eqn:nhmaxwell}) gives
\be
\label{eqn:ZBF}
 Z = \frac{\ell}{2} \left( \star_2 B - h \right),
\ee
where $\star_2$ denotes the Hodge dual with respect to the two-dimensional metric $\gamma_{ij}$ (with volume form $\eta_2$ oriented so that $d\rho \wedge \eta_2$ is the volume form of $H$).

The $\rho i$ component of equation (\ref{eqn:ricci}) gives
\be
 0 = R_{\rho i} = -\frac{1}{2} \Gamma^{-1} \gamma_{ij} \left(k^j \right)',
\ee
hence we have
\be
 k^i = {\rm constant}.
\ee
Substituting the expression (\ref{eqn:ZBF}) for $Z$ into equation (\ref{eqn:dh}) gives
\be
\label{eqn:dh1}
 \Delta'+ \frac{2\Delta \Gamma'}{\Gamma} = 0,
\ee
\be
\label{eqn:dh2}
 \left( \Gamma^{-1} k \right)' + 2 \Delta \star_2 \left( \Gamma^{-1} k \right) = - 3 \Delta B,
\ee
where, for a 1-form $\omega_i(\rho) dx^i$, we define $\omega' = \omega_i' dx^i$. Equation (\ref{eqn:dZ}) gives
\be
\label{eqn:dZ1}
 \frac{\Delta \Gamma'}{\Gamma} \star_2 1 - \Gamma^{-1} k \wedge \star_2 B = 0,
\ee
\be
\label{eqn:dZ2}
 \left( \star_2 B - \Gamma^{-1}k \right)' = \Delta B + \Delta \star_2 \left(  \Gamma^{-1} k \right) - \frac{\Gamma'}{\Gamma} \star_2 B.
\ee
Solving equation (\ref{eqn:dh1}) gives
\be
 \Delta = \frac{\Delta_0}{\Gamma^2},
\ee
where $\Delta_0$ is a non-negative constant. There are two cases to consider depending on whether $\Delta_0>0$ or $\Delta_0=0$. These correspond to non-static and static near-horizon geometries respectively.

\subsection{Non-static near-horizon geometry}

Consider first the non-static case $\Delta_0>0$. In this case, the constants $k^i$ cannot both vanish for if they did then $h$ would be closed and lemma 1 then gives $\Delta=0$, contradicting $\Delta_0>0$.

Equation (\ref{eqn:dh2}) determines $B$, which can be plugged into (\ref{eqn:ZBF}) to obtain
\be
\label{eqn:Zsol}
 Z = \frac{\ell}{2} \left[ - \frac{1}{3\Delta} \star_2 \left( \Gamma^{-1} k \right)' - \frac{1}{3} \Gamma^{-1} k + \frac{\Gamma'}{\Gamma} d\rho \right].
\ee
Substituting the expression for $B$ into (\ref{eqn:dZ1}) yields
\be
k^i(\Gamma^{-1}k_i)' = \frac{3\Delta_0^2 \Gamma'}{\Gamma^4}
\ee
and, since $k^i$ are constants, this can be integrated to give
\be
\label{eqn:ksq}
k^i k_i = C^2\Gamma- \frac{\Delta_0^2}{\Gamma^2}
\ee
where $C$ is a positive constant. Now consider equation (\ref{eqn:gradZ}). The $\rho\rho$ component gives
\be
\label{eqn:hdotZ}
 h \cdot Z + \frac{3}{\ell} = \frac{\ell}{2} \left( \frac{\Gamma''}{\Gamma} - \frac{{\Gamma'}^2}{2 \Gamma^2} \right).
\ee
The $ij$ component gives
\be\label{eqn:gradzij}
 \frac{\ell \Gamma'}{4 \Gamma} \gamma_{ij}' = - \frac{\Delta \ell \Gamma'}{4 \Gamma} \sqrt{\gamma} \epsilon_{ij} + \gamma_{ij} \left( h \cdot Z + 3/\ell \right) - Z_i \left( h + (3/\ell) Z \right)_j.
\ee
Let's deal first with the special case in which $\Gamma$ is constant. These two equations reveal that this implies $h = -(3/\ell) Z$ ($Z_i$ cannot vanish when $\Gamma'=0$ since this would imply that $Z$ vanishes but $Z$ has unit norm). All solutions with this property were obtained in \cite{GR1}. There are three possibilities: corresponding to the metric on $H$ being a homogeneous metric on the manifolds $Nil$, $SL(2,R)$ or $S^3$ respectively.

Now assume $\Gamma$ is non-constant. Multiply (\ref{eqn:gradzij}) by $\gamma^{ij}$ to get
\be
 \frac{\Gamma''}{\Gamma} = \frac{\Gamma'}{2\Gamma} \left( \log \gamma \right)'.
\ee
Since $\Gamma'$ is nonzero, we can divide by $\Gamma'$ and integrate to get
\be
\label{eqn:det}
 \sqrt{\gamma} = \beta^2 |\Gamma'|,
\ee
where $\beta$ is a positive constant.

Multiplying (\ref{eqn:gradzij}) by $k^i k^j$ leads to an equation which can be rearranged to read
\be
\label{eqn:kkij}
 \left( k \cdot Z \right)^2 = C^2 \Gamma - \frac{\Delta_0^2}{\Gamma^2} - \frac{C^2 \ell^2 {\Gamma'}^2}{4 \Gamma}.
\ee
But from equation (\ref{eqn:hdotZ}) we have
\be
\label{eqn:kdotZ}
 k \cdot Z = \frac{\ell}{2} \Gamma'' + \frac{\ell {\Gamma'}^2}{4 \Gamma} - \frac{3 \Gamma}{\ell} = \frac{1}{\ell \Gamma} \frac{dy}{d\Gamma},
\ee
where
\be
 y = \frac{\ell^2}{4} \Gamma{\Gamma'}^2 - \Gamma^3.
\ee
This implies that (\ref{eqn:kkij}) can be rewritten as
\be
 \left( \frac{dy}{d\Gamma} \right)^2 + C^2 \ell^2 y = - \Delta_0^2 \ell^2,
\ee
with solution
\be
 y = - \frac{C^2 \ell^2}{4} \left( \Gamma- \alpha_0 \right)^2 - \frac{\Delta_0^2}{C^2},
\ee
where $\alpha_0$ is a constant of integration. Hence we have
\be
\label{eqn:dGamma}
 {\Gamma'}^2 = \frac{4 P(\Gamma)}{\ell^2 \Gamma},
\ee
where
\be
 P(\Gamma) = \Gamma^3 - \frac{C^2 \ell^2}{4} \left( \Gamma- \alpha_0 \right)^2 - \frac{\Delta_0^2}{C^2}.
\ee
Note that $P(\Gamma) \ge 0$ implies that $k$ cannot vanish anywhere unless $\alpha_0^3 = \Delta_0^2/C^2$.

At this point, it is convenient to exploit the $GL(2,R)$ symmetry to set $k^1=1$, $k^2=0$. Equation (\ref{eqn:ksq}) then gives
\be
 \gamma_{11} = C^2\Gamma- \frac{\Delta_0^2}{\Gamma^2}.
\ee
Plugging equation (\ref{eqn:dGamma}) into equation (\ref{eqn:kdotZ}) and using the explicit form of $Z$ (equation (\ref{eqn:Zsol})) gives an ODE for $\gamma_{12}$:
\be
\frac{d}{d\Gamma} \left( \frac{\gamma_{12}}{\gamma_{11}} \right) = \frac{\Delta_0 \beta^2}{(C^2 \Gamma^3 - \Delta_0^2)^2} \left( 2 C^2 \Gamma^3 - 3 C^2 \alpha_0 \Gamma^2 + \Delta_0^2 \right).
\ee
This can be integrated to give
\be
 \gamma_{12} = \frac{\Delta_0 \beta^2 (\alpha_0 - \Gamma)}{C^2 \Gamma^3 - \Delta_0^2} \gamma_{11},
\ee
plus a constant times $\gamma_{11}$ which can be eliminated using the remaining $GL(2,R)$ freedom to shift $x^1$ by a constant times $x^2$. Finally we can get $\gamma_{22}$ from (\ref{eqn:det}), using the freedom to rescale $x^2$ to set $\beta=1$. The 2-metric $\gamma_{ij}$ is now fully determined in terms of $\Gamma$, so it is convenient to use $\Gamma$, instead of $\rho$ as the 3rd coordinate on $H$.
The full near-horizon solution is
\bea
\label{NHmetricA}
 g_{ab}dx^a dx^b &=& \frac{\ell^2 \Gamma d\Gamma^2}{4 P(\Gamma)} + \left( C^2 \Gamma - \frac{\Delta_0^2}{\Gamma^2} \right) \left( dx^1 + \frac{\Delta_0  (\alpha_0 - \Gamma)}{C^2 \Gamma^3 - \Delta_0^2} dx^2 \right)^2 + \frac{4 \Gamma P(\Gamma)}{\ell^2 (C^2 \Gamma^3 - \Delta_0^2)} (dx^2)^2, \nonumber \\
 \Delta &=& \frac{\Delta_0}{\Gamma^2}, \qquad k = \frac{\partial}{\partial x^1}, \qquad h = \Gamma^{-1} (k - d\Gamma),
\eea
where
\be
P(\Gamma) = \Gamma^3 - \frac{C^2 \ell^2}{4} \left( \Gamma- \alpha_0 \right)^2 - \frac{\Delta_0^2}{C^2}
\ee
with $C$ and $\Delta_0$ positive constants and $\alpha_0$ an arbitrary constant. It can be checked that the remaining equations of section \ref{sec:nheqs} are all satisfied.

To summarize, we have determined all non-static near-horizon solutions admitting two commuting Killing fields. There are several cases:
\begin{enumerate}
\renewcommand{\labelenumi}{\Roman{enumi}}
\item{If $\Gamma$ is constant then the near-horizon solution is
one of the solutions given in section 3.2 of \cite{GR1}, for which
$H$ is locally isometric to a homogeneous $SL(2,R)$, $Nil$ or
$S^3$ manifold. The latter case arises as the near-horizon limit
of the asymptotically AdS black hole solutions constructed in \cite{GR1}.} \item{If
$\Gamma$ is non-constant then the near-horizon solution is
(\ref{NHmetricA}).}
\end{enumerate}

\subsection{Static near-horizon geometry}

We now turn to the static case. This is analyzed in Appendix A,
where it is proved that the near-horizon solution must be one of
the following:
\begin{enumerate}
\renewcommand{\labelenumi}{\Roman{enumi}}
\setcounter{enumi}{2} \item{ If $\Gamma$ is constant then the
near-horizon solution is the solution given in section 3.2 of
\cite{GR1} for which $H$ is locally isometric to $R \times H^2$.
This is the near-horizon limit of the supersymmetric black
``string" of \cite{KS}. } \item{ A solution (\ref{eqn:blackring}) that can be obtained
by taking the limit $\Delta_0 \rightarrow 0$ (with other constants
and the coordinates fixed) of the non-static solution
(\ref{NHmetricA}). (This amounts to simply setting $\Delta_0=0$ in
the solution (\ref{NHmetricA}).) } \item{ A solution (\ref{eqn:noncompact}) that can be
obtained from (\ref{NHmetricA}) by setting $\Delta_0^2 = C^2
\Gamma_0^3$, $x^1 = \hat{x}^2/C$, $x^2 = C \ell^2
\hat{x}^1/4\Gamma_0^{3/2}$ and taking $C \rightarrow 0$ with
$\Gamma_0$, $\alpha_0$, $\Gamma$ and $\hat{x}^i$ fixed. \item The
$AdS_5$ solution obtained in section 3.2 of \cite{GR1}, for which
$H$ is locally isometric to hyperbolic space. }
\end{enumerate}

The near-horizon geometry of the supersymmetric black holes of
\cite{Chong:2005} is non-static with non-constant $\Gamma$ and
hence must be described by (II) which we shall prove
below. We shall also see that solution (IV) describes the
near-horizon geometry of a supersymmetric black ring but suffers from a conical singularity.

\subsection{Global considerations}

So far, the discussion has been local. However, we are interested mainly in solutions for which $H$ is compact since this is true of any solution arising as the near-horizon limit of a black hole. A compact 3-manifold admitting 
a $U(1) \times U(1)$ symmetry must be homeomorphic to $T^3$, $S^1 \times S^2$, $S^3$ or a lens space (see e.g. \cite{gowdy}). This excludes the solutions (I) and (III) with constant $\Gamma$ except for the solution of \cite{GR1} with $H$ locally isometry to $S^3$. The known black hole solutions with this near-horizon geometry \cite{GR1} arise as a special case of those of \cite{Chong:2005}, and since we shall show that the near-horizon limit of the latter solutions is described by (\ref{NHmetricA}), it follows that the $S^3$ solution with constant $\Gamma$ can be obtained as a limit of (\ref{NHmetricA}), so we shall not consider this solution further. 

The solution (VI) is excluded because $H$ cannot be compactified without breaking the rotational symmetries. So consider the other solutions with non-constant $\Gamma$. 
Note that $\gamma_{11}$ is a scalar invariant: it is the norm of $\partial/\partial x^1$ (which is some linear combination of $m_1 $ and $m_2$). Furthermore, for all of the solutions with nonconstant $\Gamma$, $\gamma_{11}$ is a monotonic function of $\Gamma$ when $\Gamma>0$. This implies that $\Gamma$ can be expressed uniquely in terms of the invariant $\gamma_{11}$ and is therefore globally defined.

If $H$ is compact and $\Gamma$ nonconstant then $\Gamma$ must achieve a distinct minimum and maximum on $H$. Hence $d\Gamma$ must vanish at two distinct (positive) values of $\Gamma$. By calculating $(d\Gamma)^2$, one finds that this is not possible for solution (V) above so this solution is necessarily noncompact. This leaves solutions (II) and (IV). Hence any compact solution with nonconstant $\Gamma$ must be described by the line element (\ref{NHmetricA}), whether static ($\Delta_0=0$) or not ($\Delta_0>0$).

\begin{figure}
\centering{\psfig{file=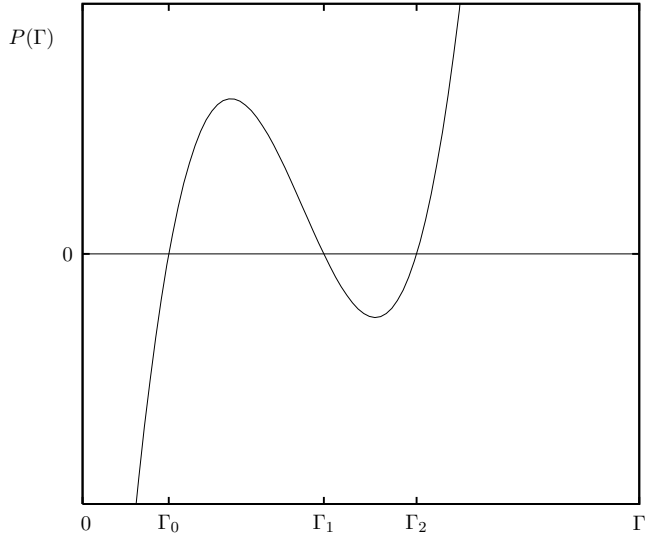,width=8.5cm}}
\begin{picture}(0,0)(180,0)
\end{picture}
\caption{Sketch of $P(\Gamma)$ corresponding to a compact $H$.  Note  $P(\Gamma) > 0$ for $\Gamma_0 < \Gamma < \Gamma_1$.}
\label{fig:cubic}
\end{figure}

For this solution we have $(d\Gamma)^2 = 4P(\Gamma)/(\ell^2 \Gamma)$, so $P(\Gamma)$ must be non-negative. Compactness implies that $P(\Gamma)$ must have two distinct positive roots $\Gamma_0$, $\Gamma_1$ with $P(\Gamma)$ positive for $\Gamma_0 < \Gamma < \Gamma_1$. It is not hard to see that this is only possible if $P(\Gamma)$ has a third root $\Gamma_2 > \Gamma_1$. (We can't have $\Gamma_2 = \Gamma_1$ since this would make the proper distance to $\Gamma=\Gamma_1$ infinite, i.e., $H$ would not be compact.) So, for a compact horizon, $P(\Gamma)$ must have the form sketched in figure~\ref{fig:cubic}. This imposes restrictions on the parameters of the solution.

At first sight, it appears that (\ref{NHmetricA}) is a 3-parameter solution. However, it is not hard to see that $\Gamma$ is defined only up to multiplication by a positive constant. This freedom can be used to eliminate one of the parameters. More explicitly, if $K$ is a positive constant then the (\ref{NHmetricA}) is invariant under the rescaling
\be
\label{scale}
\Gamma \rightarrow K \Gamma, \qquad x^{1} \rightarrow K^{-1}x^{1}, \qquad \qquad C^2 \rightarrow KC^2, \qquad \Delta_0 \rightarrow K^2 \Delta_0, \qquad \alpha_0 \rightarrow K \alpha_0.
\ee

\subsection{$\Delta_0=0$: unbalanced black ring}\label{sec:ubr}

The metric of $H$ is
\be
g_{ab}dx^a dx^b = \frac{\ell^2 \Gamma d\Gamma^2}{4 P(\Gamma)} + C^2 \Gamma ( dx^1 )^2+  \frac{P(\Gamma)}{\Gamma^2} (dx^2)^2,
\ee
where we have rescaled $x^2$. The metric has conical singularities at $\Gamma=\Gamma_0$ and $\Gamma=\Gamma_1$, where $\partial/\partial x^2$ vanishes. If one could remove these (by identifying $x^2$ with a suitable period) then one would be left with a smooth geometry of topology $S^1 \times S^2$ with the $S^1$ parameterized by $x^1$ and the $S^2$ by $\Gamma$ and $x^2$. (Note that in this case we have $m_i \propto \partial/\partial x^i$.) Thus it would describe the horizon of a supersymmetric black ring.

The necessary and sufficient condition for $P(\Gamma)$ to have distinct real positive roots is
\be
 0< \frac{\alpha_0}{C^2 \ell^2} < \frac{1}{27}.
\ee
It is convenient to define a parameter $b$ by
\be
\label{eqn:bdef}
 \frac{\alpha_0}{C^2 \ell^2} = \frac{(1-b^2)^2}{(b^2+3)^3},
\ee
where $0 < b < 1$. This defines $b$ uniquely since the function on the RHS decreases monotonically from $1/27$ at $b=0$ to $0$ at $b=1$. We can now find explicit expressions for the roots:
\be
 \Gamma_0 = \frac{\alpha_0 (b^2+3)}{4}, \qquad \Gamma_1 = \frac{\alpha_0 (b^2+3)}{(1+b)^2}, \qquad \Gamma_2 = \frac{\alpha_0 (b^2+3)}{(1-b)^2}.
\ee
The coordinate $x^2$ can be periodically identified to eliminate a conical singularity at $\Gamma=\Gamma_0$. The necessary and sufficient condition for the absence of a conical singularity of $\Gamma=\Gamma_1$ is then
\be
\label{balance}
 \left( \frac{\Gamma_1}{\Gamma_0} \right)^{3/2} = \frac{\Gamma_2 - \Gamma_1}{\Gamma_2 - \Gamma_0},
\ee
which implies $b=1$. This is not allowed since $b<1$ and hence one can never eliminate conical singularities from this metric\footnote{In fact
this is obvious since the LHS of (\ref{balance}) is greater than one and the RHS is
less than one.}. So, at best, this solution describes the near-horizon geometry of an {\it unbalanced} supersymmetric black ring: the conical singularity provides the force required to hold the ring in equilibrium.

Note that $b \rightarrow 1$ as $\ell \rightarrow \infty$, which suggests that the conical singularities {\it can} be eliminated as the cosmological constant is turned off. This is indeed true: in this limit, our $\Delta_0 = 0$ solution reduces to the $AdS_3 \times S^2$ near-horizon geometry of the supersymmetric black ring of \cite{bpsring}.\footnote{To see this explicitly, define a new coordinate $\theta$ by $\Gamma = \Gamma_0 \cos^2 \theta + \Gamma_1 \sin^2 \theta$, use equation (\ref{eqn:bdef}) to write $\ell$ in terms of $C$, $\alpha_0$ and $b$ and take the limit $b \rightarrow 1$ with $C$, $\alpha_0$ and $\theta$ fixed and rescaling $x^2$ as appropriate. Note that $\Gamma$ becomes constant in this limit.}

It is interesting to consider the five-dimensional near-horizon geometry, which, after a coordinate change $r=\Gamma R$, is
\be
 ds^2 = \Gamma \left[-C^2 R^2 dv^2 + 2 dv dR + C^2 \left(dx^1 + R dv \right)^2 \right] + \frac{\ell^2 \Gamma d\Gamma^2}{4 P(\Gamma)} + \frac{P(\Gamma)}{ \Gamma^2} (dx^2)^2,
\ee
it can be checked that the expression in the square brackets is the line element of $AdS_3$. Hence the five-dimensional near-horizon geometry is a warped product of $AdS_3$ and a deformed $S^2$ with a conical singularity at one pole.

Our solution can be oxidized on $S^5$ to give a solution of type IIB supergravity \cite{chamblin}. Viewing $S^5$ as an $S^1$ bundle over $CP^2$, the ten-dimensional solution is a warped product of $AdS_3$ with a 7-manifold $M_7$, which is an $S^1$ bundle over $S^2 \times CP^2$ where the $S^2$ is our singular $S^2$. In fact, this ten-dimensional solution has been encountered before in a general exploration of such warped products \cite{Gaunt1}. Furthermore, it has been shown that the solution {\it can} be made into a globally regular metric if one takes $M_7$ to have a different topology, namely an $S^1$ bundle over a 6-manifold $B_6$, where $B_6$ is an $S^2$ fibration over $CP^2$ \cite{Gaunt1}. The $S^1$ and $S^2$ here are not the same as for the uplift of our solution: if our Kaluza-Klein $S^1$ is parameterized by a coordinate $\psi$ (and the $S^2$ by $\Gamma$ and $x^2$) then the regular solution has $S^1$ parameterized by $x^2$ and $S^2$ parameterized by $\Gamma$ and $\psi$.

So although our near-horizon geometry is necessarily singular in five dimensions, it can be made regular if lifted to ten dimensions. However, the resulting solution can no longer be reduced to five dimensions so one loses any interpretation in terms of five-dimensional black holes. It is natural to wonder whether the regular ten-dimensional solution might describe the near-horizon geometry of a supersymmetric, asymptotically $AdS_5 \times S^5$, black hole with horizon topology $S^1 \times M_7$.

\subsection{$\Delta_0 >0$: topologically spherical black
hole}\label{sec:tsbh}

Write the metric on $H$ as
\be
\label{eqn:rhometric}
 g_{ab}dx^a dx^b =  \frac{\ell^2 \Gamma d\Gamma^2}{4 P(\Gamma)} + A(\Gamma) \left( dx^1 + \omega(\Gamma) dx^2 \right)^2 + B(\Gamma) (dx^2)^2.
\ee
There are two cases to analyze depending on whether or not $\alpha_0$ coincides with $\Gamma_0$ (note that $\alpha_0$ cannot coincide with $\Gamma_1$ since $P'(\Gamma_1)<0$ and $P'(\alpha_0)>0$). If $\alpha_0 \ne \Gamma_0$ then $A(\Gamma)=(C^2/\Gamma^2)[P(\Gamma)+(C^2\ell^2/4)(\Gamma-\alpha_0)^2]$ is positive for $\Gamma_0 \le \Gamma \le \Gamma_1$. We shall treat this case
here. The special case $\alpha_0=\Gamma_0$ is treated
separately in Appendix B. It turns out that it can be obtained as
a limit of the generic case with no further complications.

The range of $\Gamma$ is $\Gamma_0 \le \Gamma \le \Gamma_1$. The 2-metric $\gamma_{ij}$ is non-degenerate within this range but degenerates at the endpoints, where $B(\Gamma)$ vanishes. This implies that the Killing field $\omega(\Gamma_i) \partial/\partial x^1 - \partial/\partial x^2$ vanishes at $\Gamma=\Gamma_i$, $i=0,1$. Now, using the fact that $\Gamma_i$ is a root of $P(\Gamma)$, we have
\be
 \omega(\Gamma_i) = \frac{4 \Delta_0}{C^4 \ell^2 \left(\alpha_0-\Gamma_i \right) },
\ee
which implies that $\omega(\Gamma_0) \ne \omega(\Gamma_1)$. Hence the Killing field that vanishes at $\Gamma=\Gamma_0$ is distinct from the one that vanishes at $\Gamma=\Gamma_1$. In order to avoid conical singularities at $\Gamma=\Gamma_0,\Gamma_1$ we must assume that these two Killing fields have closed orbits, in other words they generate rotational symmetries. Hence our two rotational Killing fields $m_i$ must be proportional to these two Killing fields:
\be
 m_1 = -d_1 \left( \omega(\Gamma_0) \frac{\partial}{\partial x^1} -   \frac{\partial}{\partial x^2} \right), \qquad m_2 = -d_2 \left( \omega(\Gamma_1) \frac{\partial}{\partial x^1} -   \frac{\partial}{\partial x^2} \right),
\ee
for some non-zero constants $d_1,d_2$. Introducing adapted coordinates $\phi_i$ so that $m_i = \partial/\partial \phi_i$ we have
\be
 x^1 = -\left[\omega(\Gamma_0) d_1 \phi_1 + \omega(\Gamma_1) d_2 \phi_2 \right], \qquad x^2 = d_1 \phi_1 + d_2 \phi_2.
\ee The condition $\phi_i \sim \phi_i + 2\pi$ fixes the constants
$d_i$ up to signs: in order to avoid conical singularities one must
take
\begin{equation}
 |d_1| =  \frac{\ell^2 (C^2\Gamma_0^3 -
 \Delta_0^2)^{\frac{1}{2}}}{2P'(\Gamma_0)}, \qquad |d_2| = \frac{\ell^2
 (C^2\Gamma_1^3 - \Delta_0^2)^{\frac{1}{2}}}{2|P'(\Gamma_1)|}.
\end{equation}
The solution is now globally regular. It is clear that $H$ has $S^3$ topology,\footnote{
More precisely, the covering space of $H$ has $S^3$ topology. $H$ could actually be a lens space if additional identifications are made.} with $m_1$ vanishing at $\Gamma=\Gamma_0$ and $m_2$ vanishing at $\Gamma=\Gamma_1$. In Appendix B it is shown that the coordinate change from $(x^1,x^2)$ to $(\phi_1,\phi_2)$ is also valid for the special case $\alpha_0 = \Gamma_0$ (although $d_1 \rightarrow 0$ as $\alpha_0 \rightarrow \Gamma_0$, the product $d_1 \omega(\Gamma_0)$ remains nonzero in this limit). Hence we do not need to treat this case separately any longer.

We shall now show that the near-horizon solution we have obtained is locally isometric to that of the supersymmetric black hole solution of Chong {\it et al} \cite{Chong:2005}. It is convenient to use the roots $\Gamma_{i}$ as parameters, as opposed
to $(C^2,\alpha_{0},\Delta_{0})$. These are related by
\begin{eqnarray}
\label{eqn:constab} C^2 &=& \frac{4}{\ell^2} (\Gamma_{0} +
\Gamma_{1} + \Gamma_{2}) \qquad \alpha_{0} =
\frac{\Gamma_{0}\Gamma_{1} + \Gamma_{0}\Gamma_{2} +
\Gamma_{1}\Gamma_{2}}{2(\Gamma_{0} + \Gamma_{1} + \Gamma_{2})}
\nonumber \\ \Delta^2_{0} &=& C^2 \Gamma_{0}\Gamma_{1}\Gamma_{2} -
\frac{C^4 \ell^2 \alpha_{0}^2}{4} .
\end{eqnarray}
Note that $\Gamma_i$ are not totally arbitrary positive numbers: they are constrained by $\Delta_0^2>0$.
Use the scale transformations (\ref{scale}) on $\Gamma$ to fix
\begin{equation}
\frac{1}{\Gamma_{0}} + \frac{1}{\Gamma_{1}} - \frac{1}{\Gamma_{2}} = 2
\end{equation} which can always be done (i.e. the corresponding $K>0$) since $0<\Gamma_0<\Gamma_1<\Gamma_2$.
We can define two constants $a,b$ so that the roots
are parameterized as:
\begin{equation}
\label{eqn:abdef}
\Gamma_{0} = \frac{1}{1+ag} \qquad \Gamma_{1} = \frac{1}{1+bg} \qquad \Gamma_{2} = \frac{1}{g(a+b)},
\end{equation}
where $g \equiv 1/\ell$. We need $\Gamma_i$ to be positive and
correctly ordered, which gives the restrictions
\be
 g^{-1} > a > |b|.
\ee
Note that $b$ may be negative. One can solve for $C$, $\alpha_0$ and $\Delta_0$ using (\ref{eqn:constab}). One finds that
\be
 \Delta_0^2 = \frac{4(ag+bg+abg^2)}{(1+ag)^2(1+bg)^2(a+b)^2},
\ee so the constraint coming from $\Delta_0^2 > 0$ is, in the notation of \cite{Chong:2005},
\be
r_0^2 \equiv g^{-1}(a+b)+ab > 0.
\ee 
We now define a coordinate $\theta$ by \be
 \Gamma = \frac{g \rho(\theta)^2}{(a+b)(1+ag)(1+bg)}
\ee
where
\be
 \rho(\theta)^2 = \frac{a+b}{g} + ab + a^2 \cos^2 \theta + b^2 \sin^2 \theta.
\ee It is easy to check that taking $0 \le \theta \le \pi/2$ does
indeed uniquely parameterize the range $\Gamma_0 \le \Gamma \le
\Gamma_1$ (although $\theta=0$ corresponds to $\Gamma=\Gamma_1$ and
$\theta=\pi/2$ to $\Gamma=\Gamma_0$). We then have \be
 \cos^2 \theta = \frac{\Gamma-\Gamma_0}{g^2(a^2-b^2) \Gamma_0 \Gamma_1 \Gamma_2}, \qquad \sin^2 \theta = \frac{\Gamma_1 - \Gamma}{g^2(a^2-b^2) \Gamma_0 \Gamma_1 \Gamma_2},
\ee
\be
 P(\Gamma) = \frac{g(a-b)^2}{(1+ag)^3(1+bg)^3 (a+b)} \cos^2 \theta \sin^2 \theta \Delta_\theta,
\ee
where
\be
  \Delta_\theta \equiv 1 - a^2 g^2 \cos^2 \theta - b^2 g^2 \sin^2 \theta,
\ee
and
\be
 \frac{\ell^2 \Gamma d\Gamma^2}{4 P(\Gamma)} = \frac{\rho(\theta)^2 d\theta^2}{\Delta_\theta}.
\ee
For completeness, we give the expressions for the constants $d_i$:
\bea d_1 = \frac{a+2b+2abg+gb^2}{g(1-bg)(a-b)},
\qquad d_2 = -\frac{b+2a+2abg+ga^2}{g(1-ag)(a-b)} \eea
We have checked that our 3-metric in the $(\theta, \phi_1,
\phi_2 )$ coordinates agrees exactly with the horizon metric of
Chong {\it et al} \, in their $(\theta, \psi, \phi)$
coordinates\footnote{Actually before taking the near horizon limit
of their metric one needs to introduce coordinates which are valid
on the horizon. In particular the angles $\psi$ and $\phi$ must be shifted by functions of $r$ (see the analysis in~\cite{KLR}) and
it is these new angular variables which we are referring to.} with
$\phi_1={\psi}$ and $\phi_2={\phi}$. 

It is interesting to note that the near-horizon geometry is considerably simpler in the coordinates $(\Gamma,x^1,x^2)$ than in the coordinates $(\theta,\phi_1,\phi_2)$. 

\section{Discussion}

In this paper we have determined the most general regular, supersymmetric,
near-horizon geometry of an asymptotically $AdS_5$ black hole solution of minimal gauged supergravity which admits two rotational isometries. 
We found that the only such solution is the near-horizon geometry of 
the topologically spherical supersymmetric black hole discovered by Chong {\it et al} \cite{Chong:2005}. Hence if new supersymmetric black hole solutions exist then either they do not have two rotational symmetries, or they have the same near-horizon geometry as the known solutions.

Our result implies that exotic topologies for supersymmetric black holes (such
as black rings) are not allowed, unless they possess fewer than two
rotational symmetries. We should emphasise that no known black hole in
five dimensions possesses fewer than two rotational symmetries, including
the asymptotically flat supersymmetric black rings of ungauged
supergravity \cite{bpsring}. It has been conjectured, however, that there could exist asymptotically flat black holes with only one rotational symmetry \cite{Reall:NH}, essentially because this is all one is guaranteed in general \cite{HIW:2005}.

Curiously we did find a near-horizon geometry which describes a
black ring, however it necessarily possesses a conical singularity. Physically this can be interpreted as suggesting that supersymmetric black rings in AdS cannot be "balanced", i.e., rotation and electromagnetic repulsion is not enough to counterbalance gravitational attraction (which is enhanced in AdS). However, known unbalanced ring solutions (e.g. \cite{vacring,dipole}) can be balanced by increasing the angular momentum, so it seems likely the same will be true here (the mass and/or charge would also have to change for consistency with the BPS inequality). Hence
our singular near-horizon geometry may be interpreted as evidence in favour of the existence of regular non-supersymmetric anti-de Sitter black rings.

It is interesting to compare our result with the corresponding result for the ungauged theory \cite{Reall:NH}. In the latter theory, one obtains a complete classification assuming only supersymmetry and compactness of the horizon. The allowed near-horizon geometries all admit two rotational symmetries (in fact $H$ is homogeneous), but this is an output, not an input. In the gauged theory, we had to input the rotational symmetries as an assumption. It would be nice if this assumption could be relaxed and the general solution of the equations of section \ref{sec:nheqs} determined (for compact $H$).

In the ungauged theory it is possible to construct a global uniqueness argument for supersymmetric black holes of spherical topology~\cite{Reall:NH}. It is obviously desirable to have a corresponding result for the gauged theory, i.e., prove that the solution of \cite{Chong:2005} is the only supersymmetric asymptotically AdS black hole solution, even assuming two rotational symmetries. However, the arguments of \cite{Reall:NH} rely heavily on features of the ungauged theory that do not extend to the gauged theory so new ideas would be required to do this.

In the ungauged theory, it is straightforward to extend the argument to allow for abelian vector multiplets \cite{jan} and the same is probably true in the gauged theory. So it seems likely that any supersymmetric, asymptotically AdS black hole solution of gauged supergravity coupled to abelian vector multiplets must have the same near-horizon geometry as the solution of \cite{KLR} if it admits two rotational symmetries. 

\section*{Acknowledgments}
HKK and JL are supported by PPARC. HSR is a Royal Society University
Research Fellow. We are grateful to Sigbjorn Hervik for helpful correspondence.

\appendix

\section{Static near-horizon geometry: $\Delta_0 = 0$}
The analysis can be divided into three subcases.\\

\noindent (I) First, consider $k \equiv 0$ and $B \equiv 0$. Then $h=-(2/\ell) Z$. It was shown in section 3.2 of \cite{GR1} that this implies that the solution is just $AdS_5$ with vanishing Maxwell field. $H$ is locally isometric to hyperbolic space, and hence cannot be compactified without breaking the rotational symmetries.\\

\noindent (II) Consider the case $k \equiv 0$ and $B$ not
identically zero. Equation (\ref{eqn:dZ2}) implies \be
 \star_2 B = \Gamma^{-1} \omega,
\ee where $\omega_i$ are constants (note $\omega_i=0$ is covered by
case (I) and thus we will assume this is not the case here). We have
\be
 h = -\frac{\Gamma'}{\Gamma} d\rho, \qquad Z = \frac{\ell}{2} \left( \Gamma^{-1}   \omega + \frac{\Gamma'}{\Gamma} d\rho \right).
\ee  The $\rho\rho$ component of (\ref{eqn:gradZ}) gives
\be
 \frac{\Gamma''}{\Gamma} + \frac{\Gamma'^2}{2 \Gamma^2} = \frac{6}{\ell^2},
\ee
which implies that $\Gamma'$ is not identically zero hence we can change variable from $\rho$ to $\Gamma$ and rewrite this equation as
\be
 \frac{d}{d\Gamma} \left( \Gamma \Gamma'^2 - \frac{4\Gamma^3}{\ell^2} \right) =0 ,
\ee
with solution
\be
 \Gamma'^2 = \frac{4}{\ell^2 \Gamma}P(\Gamma),
\ee
where
\be
 P(\Gamma) = \Gamma^3 -\Gamma_0^3,
\ee and for convenience the integration constant has been chosen
such that $\Gamma_0$ is the real root of $P(\Gamma)$. Using the fact
that $Z$ is a unit one-form one can deduce that: \be
\label{normomega} \omega^i\omega_i = \frac{4\Gamma_0^3}{\ell^2
\Gamma} \ee and thus $\Gamma_0> 0$. The $ij$ component of
(\ref{eqn:gradZ}) reduces to \be \label{eqn:dgammaij}\frac{\ell
\Gamma'}{4\Gamma} {\gamma_{ij}}' = \frac{1}{\ell} \left( 1+
\frac{2\Gamma_0^3}{\Gamma^3} \right) \gamma_{ij} - \frac{3\ell}{4
\Gamma^2} \omega_i \omega_j. \ee We can use the $GL(2,R)$ freedom to
set $\omega_1=1$ and $\omega_2=0$, i.e. $\omega =dx^1$. Therefore
$\gamma_{12}$ and $\gamma_{22}$ satisfy the same first order ODE
which can be written as: \be \frac{d}{d\Gamma} \log \gamma_{i2} =
\frac{P'(\Gamma)}{P(\Gamma)} - \frac{2}{\Gamma}. \ee Thus
$\gamma_{12}$ and $\gamma_{22}$ are both equal to some constant
times $P(\Gamma)/\Gamma^2$. This means that we can use the remaining
$GL(2,R)$ freedom to set $\gamma_{12}=0$ and $\gamma_{22} =
P(\Gamma)/\Gamma^2$. Multiplying (\ref{eqn:dgammaij}) by
$\gamma^{ij}$ implies \be \frac{d}{d\Gamma} \log \gamma =
\frac{P'(\Gamma)}{P(\Gamma)} -\frac{1}{\Gamma} \ee where $\gamma$ is
the determinant of $\gamma_{ij}$, and hence $\gamma= C^2
P(\Gamma)/\Gamma$ where $C$ is a positive constant. These results
imply $\gamma_{11}= C^2\Gamma$. Note that
$\omega_i\omega^i=\gamma^{11}$ and thus upon comparison with
(\ref{normomega}) we deduce $C^2=\ell^2/4\Gamma_0^3$. The rest of
the equations in section \ref{sec:nheqs} are satisfied without further constraint. Using
$\Gamma$ instead of $\rho$ as a coordinate allows one to write the
metric on the horizon as: \be 
\label{eqn:noncompact}
g_{ab}dx^adx^b = \frac{\ell^2 \Gamma
d\Gamma^2}{4 P(\Gamma)} + \frac{\ell^2}{4\Gamma_0^3} \Gamma (dx^1)^2
+ \frac{P(\Gamma)}{\Gamma^2} (dx^2)^2, \ee This metric is smooth for
$\Gamma
>\Gamma_0$ (recall necessarily we have $\Gamma_0>0$). There is a conical singularity at $\Gamma= \Gamma_0$ but
this can be removed by appropriately identifying $x^2$. The
3-manifold this defines is topologically $R^2\times S^1$. By
rescaling $x^1$ suitably and letting
$\Gamma_0 \to 0$ we recover case (I).\\

\noindent (III) Consider the case where $k$ is not identically zero.
Equation (\ref{eqn:dh2}) implies that $\Gamma^{-1} k_i = \const$,
and therefore one can again deduce from (\ref{eqn:dZ2}) that: \be
\star_2 B =\Gamma^{-1} \omega \ee where $\omega_i$ are constants.
Expanding $\star_2B$ in the basis $\{ k, \star_2 k \} $ and using
(\ref{eqn:dZ1}) implies \be \omega = \alpha_0 \Gamma^{-1} k \ee for
some constant $\alpha_0$. Therefore \be Z= \frac{\ell}{2}\left[
\left( \frac{\alpha_0}{\Gamma} -1 \right) \Gamma^{-1}k +
\frac{\Gamma'}{\Gamma} d\rho \right]. \ee Note that if $\Gamma$ is a
constant, $h=\alpha Z$ for some constant $\alpha$. This case was
considered in section 3.2 of~\cite{GR1}, where it was proved that
$H$ must be locally isometric to $R \times H^2$.  In the following
we assume $\Gamma$ is non-constant. From the fact that $Z$ is a unit
one-form one can show that: \be {\Gamma'}^2 = \frac{4
P(\Gamma)}{\ell^2 \Gamma} \ee where \be \label{eqn:Pring} P(\Gamma)
= \Gamma^3 - \frac{C^2 \ell^2}{4} (\Gamma-\alpha_0)^2 \ee and we
have defined the constant $C>0$ by $C^2 = \Gamma^{-1}k_i k^i$ (since
$k^i$ are constants by assumption). Let us use the $GL(2,R)$ freedom
available to set $k^1=1$ and $k^2=0$. Using the definition of the
constant $C$ then gives $\gamma_{11}= C^2\Gamma$. Also, since
$\Gamma^{-1} k_i$ is a constant we deduce that $\gamma_{12} =
c\Gamma$ for some constant $c$. Using the $GL(2,R)$ transformations
(which leave $k^i$ invariant), we can arrange to have
$\gamma_{12}=0$. Then, the $22$ component of equation
$(\ref{eqn:gradZ})$ simplies to: \be \frac{d}{d\Gamma}\log
\gamma_{22} = \frac{P'(\Gamma)}{P(\Gamma)} - \frac{2}{\Gamma} \ee
which implies $\gamma_{22} = P(\Gamma)\Gamma^{-2}$ (the integration
constant have been fixed using the remaining $GL(2,R)$ freedom).
Collecting these results gives the horizon geometry \be
\label{eqn:blackring}
 g_{ab}dx^adx^b = \frac{\ell^2 \Gamma d\Gamma^2}{4 P(\Gamma)} +
C^2 \Gamma (dx^1)^2 + \frac{P(\Gamma)}{\Gamma^2} (dx^2)^2
\ee
with $P(\Gamma)$ given by (\ref{eqn:Pring}). The rest of the equations in section \ref{sec:nheqs} are now satisfied without further constraint. This metric is analysed in  section \ref{sec:ubr} of the main text.

\section{Special case $\alpha_0=\Gamma_0$}

This special case needs to be considered separately. The metric on $H$ still takes the form (\ref{eqn:rhometric}). For $\Gamma_0
\leq \Gamma \leq \Gamma_1$ the function $A(\Gamma)$ now vanishes for
$\Gamma=\Gamma_0$ and is positive otherwise. The function
$B(\Gamma)$ now vanishes only at one end point $\Gamma=\Gamma_1$ and
is positive in the rest of the interval.
Therefore the 2-metric $\gamma_{ij}$ is non-degenerate for
$\Gamma_0<\Gamma<\Gamma_1$ and degenerates at $\Gamma=\Gamma_0$ and
$\Gamma=\Gamma_1$. The Killing field $\partial/\partial x^1$
vanishes at $\Gamma=\Gamma_0$ and the Killing field
$\omega(\Gamma_1)\partial/\partial x^1 - \partial /
\partial x^2$ vanishes at $\Gamma=\Gamma_1$. In order to avoid
conical singularities at $\Gamma=\Gamma_0,\Gamma_1$ these two
Killing fields must have closed orbits. Thus these two Killing
fields must be proportional to $m_i$, say \be m_1 = c_1
\frac{\partial}{\partial x^1}, \qquad m_2= -d_2 \left(
\omega(\Gamma_1) \frac{\partial}{\partial x^1}
-\frac{\partial}{\partial x^2} \right). \ee Now introduce adapted
coordinates such that $m_i =
\partial / \partial \phi_i$:
\be x^1 = c_1 \phi_1-d_2 \omega(\Gamma_1) \phi_2, \qquad x^2 =d_2
\phi_2. \ee The condition $\phi_i \sim \phi_i +2\pi$ implies that in
order to avoid the conical singularities we must take: \be c_1^2 =
\frac{\ell^2}{9C^2 \Gamma_0}, \qquad d_2^2 = \frac{C^4 \ell^6
(\Gamma_1-\Gamma_0)^2}{16 P'(\Gamma_1)^2}. \ee This solution is now
globally regular and has $S^3$ topology with $m_1$ vanishing at
$\Gamma=\Gamma_0$ and $m_2$ vanishing at $\Gamma=\Gamma_1$. The
coordinate change $(x^1,x^2) \to (\phi_1,\phi_2)$ can be obtained
from the $\alpha_0 \neq \Gamma_0$ case studied in section
\ref{sec:tsbh} by taking the limit $\alpha_0 \to \Gamma_0$.

\end{document}